\documentclass[twoside,reqno]{H}
\usepackage{epsfig,cite,colordvi}
\usepackage{graphicx}
\usepackage{graphics}
\usepackage{url}
\usepackage{amssymb,amsmath,amscd,epsf}
\usepackage{times}
\usepackage{makeidx}
\usepackage[english]{babel}
\newcommand{\inieq}{\begin{eqnarray}}            
\newcommand{\fineq}{\end{eqnarray}}            

\newcommand{\eep} { $(e,e^{\,\prime}p)$ } 
\def\ee{\mbox{$\left(e,e^{\prime}\right)$\ }}

\begin{document}

\title{Recent Results of the Relativistic Green's Function Model in 
Quasielastic Neutrino and Antineutrino-Nucleus Scattering}

\runningheads{Recent Results of the Relativistic Green's Function Model in 
Quasielastic Neutrino and Antinueutrino-Nucleus Scattering}{C. Giusti, A. Meucci}

\begin{start}

\author{C. Giusti}{1}, \coauthor{A. Meucci}{1},

\index{Giusti, Carlotta}
\index{Meucci, Andrea}

\address{Dipartimento di Fisica, 
Universit\`{a} degli Studi di Pavia and \\
INFN, Sezione di Pavia, via Bassi 6 I-27100 Pavia, Italy}{1}

\begin{Abstract}
The analysis of quasielastic neutrino and antineutrino-nucleus scattering 
cross sections requires relativistic theoretical descriptions 
also accounting for the role of final-state interactions (FSI). 
In the relativistic Green's function (RGF) model FSI are described by a 
complex optical potential where the imaginary part recovers the contribution of
final-state channels that are not included in other models based on the 
impulse approximation. The RGF results are compared with the data recently 
published by the MiniBooNE and MINER$\nu$A Collaborations. 
The model is in general able to give a good description of the  
data. 
\end{Abstract}
\end{start}

\section{Introduction}

Since many years electron scattering has been devised as a 
powerful and
preferential tool to investigate nuclear properties. Several decades of
experimental and theoretical work 
have provided 
complete and detailed information on nuclear structure and interaction 
mechanisms~\cite{rep,book,bds}.
  
Additional information on nuclear properties is in principle 
available from neutrino-nucleus scattering. The main aim of neutrino
experiments, however,  is not to study nuclear properties, but to determine
neutrino properties. Neutrino physics is of great interest and involves many
different phenomena. In neutrino oscillation experiments nuclei are used 
as neutrino detectors providing relatively large cross sections. Because of 
the interest in oscillation measurements, various experimental neutrino-nucleus 
cross sections have been
presented~\cite{miniboone,miniboonenc,miniboonean,miniboonencan,sciboone,argo,minerva1,minerva2} and are planned in the near 
future.  
A proper interpretation of the experimental data requires accurate 
evaluations of neutrino-nucleus cross sections where all 
nuclear effects are well under control. 

Models developed for electron scattering and tested in comparison with 
electron scattering data have been applied to describe nuclear effects in neutrino-nucleus 
scattering. Although different, the two situations present many similar aspects and the extension to neutrino scattering of the electron scattering formalism 
is straightforward. Moreover, the large amount of detailed experimental 
information available from electron scattering makes the comparison with 
electron scattering data the best and most obvious test of the predictive power 
of a nuclear model. 

In this contribution we review recent results obtained with the 
relativistic Green's function (RGF) model. The model was originally developed,
within a nonrelativistic~\cite{eenr,eesym} and then a relativistic 
framework~\cite{ee,eea}, to describe FSI in the 
inclusive quasielastic (QE) electron scattering, where it has been widely and 
successfully tested in comparison with 
data~\cite{eenr,eesym,confee,eeex,rila13}.
But for some complications due to the Dirac matrix structure,  the formalism 
follows within both frameworks the same steps and approximations. 
The relativistic version of the model (RGF) has been extended to QE 
neutrino-nucleus 
scattering~\cite{cc,acta,acta1,confcc,compmini,prd,prd1,prd2,compnc,prd3,prd4}. 
The energy region explored in most neutrino 
experiments requires a relativistic model, where not only relativistic 
kinematics  is considered, but also nuclear dynamics and current operators are 
described within a relativistic framework. 

\section{Final-State Interactions in Quasielatic Lepton-Nucleus Scattering}

In the QE kinematic region the nuclear response to an electroweak probe is
dominated by one-nucleon processes, where the scattering occurs with only one
nucleon, which is then emitted by a direct knockout mechanism, and the remaining
nucleons of the target behave as spectators. In electron scattering experiments
the emitted nucleon can be detected in coincidence with the scattered electron.
Kinematic situations can be envisaged where the residual nucleus is left in a 
discrete eigenstate and the final state is completely determined. This is the 
exclusive one-nucleon knockout. The exclusive \eep reaction has been widely 
investigated~\cite{book}.
If only the scattered electron is detected, the final nuclear state is not determined
and the measured cross section includes all the available final states. This is
the inclusive \ee scattering.

Electron and neutrino-nucleus scattering are usually described in the 
one-boson exchange approximation, where the cross section is obtained from 
the contraction between the lepton tensor, which 
essentially depends only on the lepton 
kinematics,  and the hadron tensor $W^{\mu\nu}$, whose components 
are given by products of the matrix elements of the nuclear current between 
the initial and final nuclear states. Different models have been developed to
calculate the hadron  tensor in the exclusive and inclusive processes.

Within the QE kinematic domain, electron scattering can adequately be 
described by a model based on the nonrelativistic or relativistic impulse 
approximation (IA or RIA). For the exclusive scattering the IA 
assumes that the interaction occurs through a one-body current only with a 
quasi-free nucleon which is then knocked out of the nucleus. For the inclusive
scattering, the IA assumes that the cross section is given by the
incoherent sum of one-nucleon knockout processes due to the interaction of the
probe with all the nucleons of the nucleus. With this assumption, we have the
problem to describe the FSI between the emitted nucleon and the residual nucleus.

In the exclusive \eep reaction FSI are usually described in the distorted-wave 
impulse approximation (DWIA) by a complex optical potential (OP) where the imaginary
part gives an absorption that reduces the calculated cross section. This
reduction is essential to reproduce \eep data. Models based on a nonrelativistic
DWIA or a relativistic RDWIA are indeed able to give an excellent description 
of \eep data~\cite{rep,book,ud93,meucci01,meucci01a,epja,Giusti11}. 

In the inclusive scattering a model based on the DWIA orthe  RDWIA, where the 
cross section is given by the sum of all integrated one-nucleon knockout 
processes and FSI are described by a complex OP with an 
imaginary absorptive part, is conceptually wrong. The OP describes elastic
nucleon-nucleus and its imaginary part accounts for the fact that, if other 
channels are open besides the elastic one, part of the incident flux is lost in 
the elastically scattered beam and goes to the inelastic channels which are 
open. In the exclusive reaction, where only one channel is considered, it 
is correct to account for the flux lost in the selected channel.  In the 
inclusive scattering all the final-state channels are included, the flux lost 
in a channel must be recovered in the other channels, and in the sum over all 
the channels the flux can be redistributed but must be conserved. The DWIA 
and RDWIA do not conserve the flux. 

Different relativistic approaches have been adopted, within the RIA, to 
describe FSI in the inclusive QE electron scattering. In the relativistic 
plane-wave impulse approximation (RPWIA) the plane-wave approximation is 
assumed for the emitted nucleon wave function and FSI are neglected. 
In other models FSI are incorporated in the emitted nucleon states by using 
real potentials, either retaining only the real part of the relativistic OP 
(rROP) or using the same relativistic energy-independent mean-field 
potential that describes the initial nucleon state 
(RMF)~\cite{confee,Chiara03,cab,cab1}.  
Both the rROP and RMF conserve the flux, but the rROP is
conceptually wrong because the optical potential has to be complex, owing to 
the presence of open inelastic channels. 
Its real and imaginary parts  
are related by dispersion relations and a model that arbitrarily omits a part 
is conceptually wrong. 
We note that the RMF fulfills the dispersion relations 
and maintains the continuity equation. 
 
In the RGF model FSI are described in the inclusive scattering consistently 
with the exclusive scattering by the same complex ROP, but in the inclusive 
scattering the imaginary part conserves and redistributes the flux in all the 
channels.
Detailed discussions of the model and of its formalism can be found, 
{\textit{e.g.}}, in~\cite{eenr,eeex,ee}. Here we summarize only the main steps.

With suitable approximations, which are essentially related to the RIA, the 
components of the hadron tensor can be written in terms of the single-particle
(s.p.) optical model Green's function. The explicit calculation of the s.p. 
Green's function can be avoided exploiting its spectral 
representation, which is based on a biorthogonal expansion in terms of the 
eigenfunctions of the non-Hermitian ROP and of its Hermitian conjugate.  
The s.p. expression of the hadron-tensor 
components is then obtained in a form  which contains matrix elements of the 
same type as the RDWIA ones of the exclusive \eep process, but these matrix 
elements now involve eigenfunctions of the
ROP and of its Hermitian conjugate, where the opposite sign
of the imaginary part gives, in one case, an absorption and,
in the other case, a gain of strength. Therefore, in the RGF model the 
imaginary part of the ROP  redistributes the flux lost in the channel in the
other (inelastic) channels and in the sum over all the channels the total
flux is conserved. With the use of a complex ROP the model can recover the 
contribution of inelastic channels which are not included in other models 
based on the RIA: all the available final-state channels are included in 
the RGF and not only direct one-nucleon emission processes.

In the usual applications of the model the matrix elements are calculated using the same phenomenological 
bound and scattering states adopted in RDWIA calculations.
We note that  the use of a phenomenological ROP in the RGF calculations does 
not allow us to disentangle the contribution of a specific inelastic channel.
Moreover, different parameterizations of the phenomenological ROP are available, 
which may provide equivalently good descriptions of elastic nucleon-nucleus 
scattering data, but which can give different results in RGF calculations and 
therefore introduce theoretical uncertainties in the numerical predictions of 
the model.

The results of different relativistic descriptions of FSI, in particular of 
the RGF and RMF, have been compared in~\cite{confee}.  
The RGF and RMF results are always different from the results of the simpler
RPWIA and rROP approaches. The differences between the RMF and RGF results 
increase with the momentum transfer. The differences of the RGF results 
obtained with different parameterizations of the ROP depend on kinematics and 
are basically due to the differences in the imaginary part of the ROPs. 
The real terms are very similar and the rROP cross sections are not sensitive 
to the parameterization used in the calculations.

The RGF provides in many cases a good description of the experimental 
$\ee$ cross sections in the QE region, in 
particular in kinematic situations where the longitudinal response is 
dominant~\cite{ee,confee,eeex,rila13}. 

In the case of charged-current QE (CCQE) (anti)neutrino-nucleus scattering the situation where only the final lepton is 
detected can be treated with the same models used for the inclusive QE \ee 
scattering. The RGF and RMF results have been compared in~\cite{confcc} for 
CCQE and in~\cite{compnc} for neutral-current elastic (NCE) scattering. 
Some differences are obtained with 
respect to the corresponding \ee results. The numerical differences between 
the RGF results for electron and neutrino scattering can mainly be ascribed to 
the combined effects of the weak current, in particular its axial term, and 
the imaginary part of the ROP~\cite{confcc}. 
The differences between the RGF and RMF results can be ascribed to the 
inelastic contributions which are incorporated in the RGF 
but not in the RMF (and in other models based on the RIA).
The RGF and RMF  give in many cases close predictions, usually
different from those of the simpler RPWIA and rROP, but there are also
kinematics situations where the differences are large~\cite{confee,confcc}. 
We recall that the RMF uses as input the real, strong, energy-independent,
relativistic mean field potential that reproduces the saturation properties of 
nuclear matter and of the ground state of the nuclei involved. As such, it 
includes only purely nucleonic contribution and does not incorporate any 
information from scattering reactions. 
In contrast, the RGF uses as input complex phenomenological 
energy-dependent ROPs, which have been obtained through a fit 
to elastic proton-nucleus scattering data. Therefore, the RGF  
incorporates information from scattering reactions and takes into account 
all the allowed final states, as the s.p. Green's function contains the full 
propagator. 

Since the imaginary part of the ROP includes the overall effect of the 
inelastic channels, 
we can expect that the differences between the RGF and RMF results increase 
with the relevance of such inelastic contributions. 
The comparison of the results of different models can therefore be useful 
to evaluate the relevance of inelastic contributions. 
In the comparison with data, we may expect that the RGF can give a better 
description of those experimental cross sections which receive significant
contributions from non-nucleonic excitations and multi-nucleon processes.
This is expected to be the case \cite{compmini,Tina10} of  MiniBooNE CCQE data.

The energy dependence of the ROP, which reflects the different contribution 
of the inelastic channels which are open at different energies, makes the RGF 
results very sensitive to the kinematic conditions of the calculations. 
Different kinematic conditions are generally considered in electron and 
neutrino scattering experiments. 
While in electron-scattering experiments the beam energy is known and the cross
sections are given as a function of the energy transfer $\omega$, in neutrino 
experiments $\omega$ and the momentum transfer $q$ are not known, and the 
calculations for the comparison with data are carried out over the entire 
energy range which is relevant for the neutrino flux. 
The flux-average procedure can include contributions from different kinematic 
regions where the neutrino flux has significant strength and processes other
than direct one-nucleon emission can be important. Part of 
these contributions are recovered in the RGF  by the imaginary part of 
the ROP.

\section{Comparison with CCQE and NCE  data}

The first measurements of the CCQE
flux-averaged double-differential $\nu_{\mu} (\bar{\nu_{\mu}})$ cross
section on $^{12}$C in the few GeV region by the MiniBooNE 
Collaboration have been reported in~\cite{miniboone,miniboonean}. 
The fact that the experimental cross sections are usually 
underestimated by models based on the RIA have suggested that effects beyond 
the RIA may play a significant role at MiniBooNE 
kinematics~\cite{Tina10,Tina09,Martini,Ank,Fer,Nieves}.

The results of different descriptions of FSI have been compared 
with the MiniBooNE data in~\cite{compmini,prd}. The RPWIA, rROP, and RMF 
generally underpredict the data, only the RGF gives larger cross sections, in reasonable
agreement with the data, both for $\nu_{\mu}$ and  
$\bar{\nu_{\mu}}$ scattering.

\begin{figure}[t]
\centering
\includegraphics[scale=0.27]{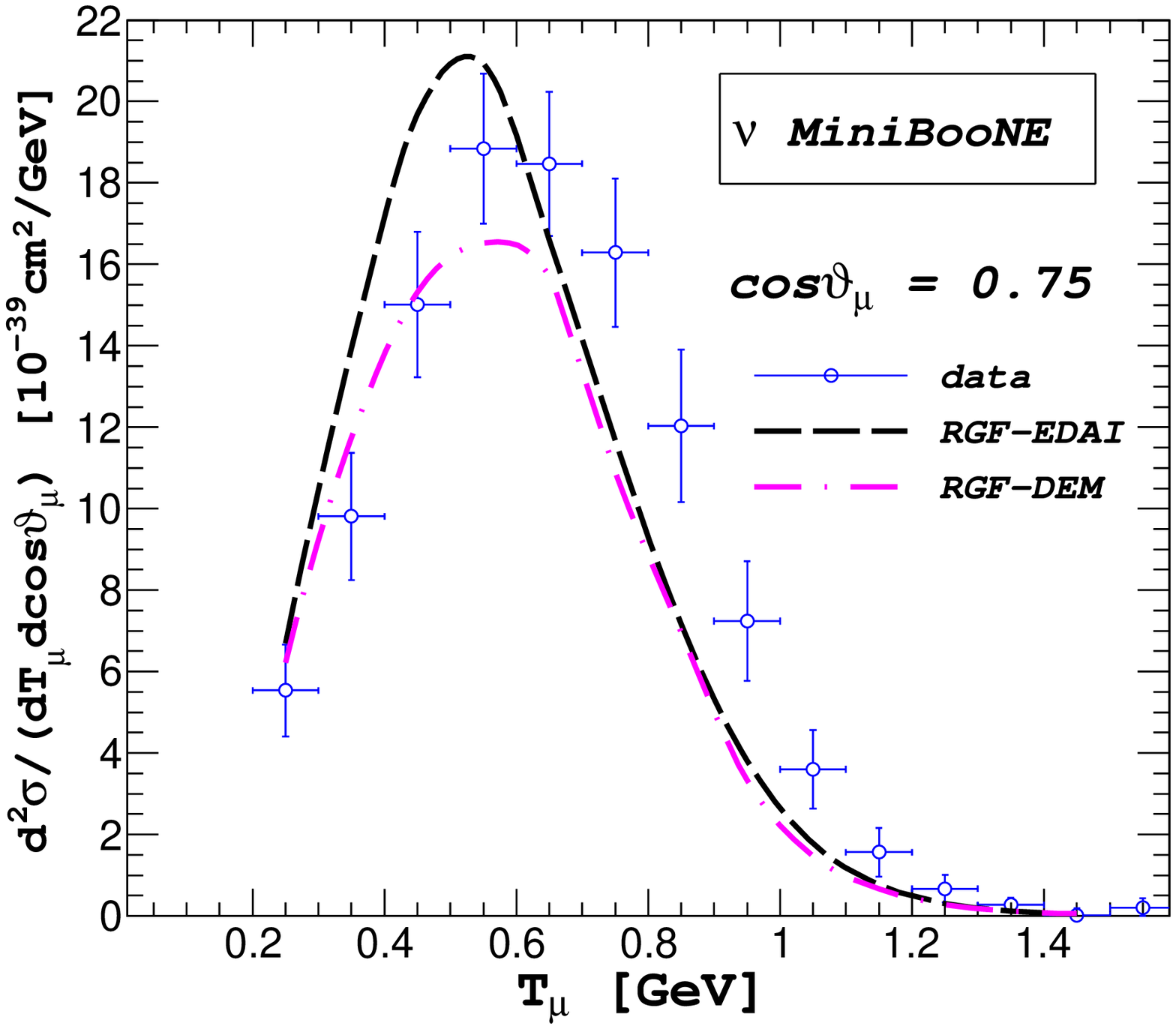} 
\includegraphics[scale=0.27]{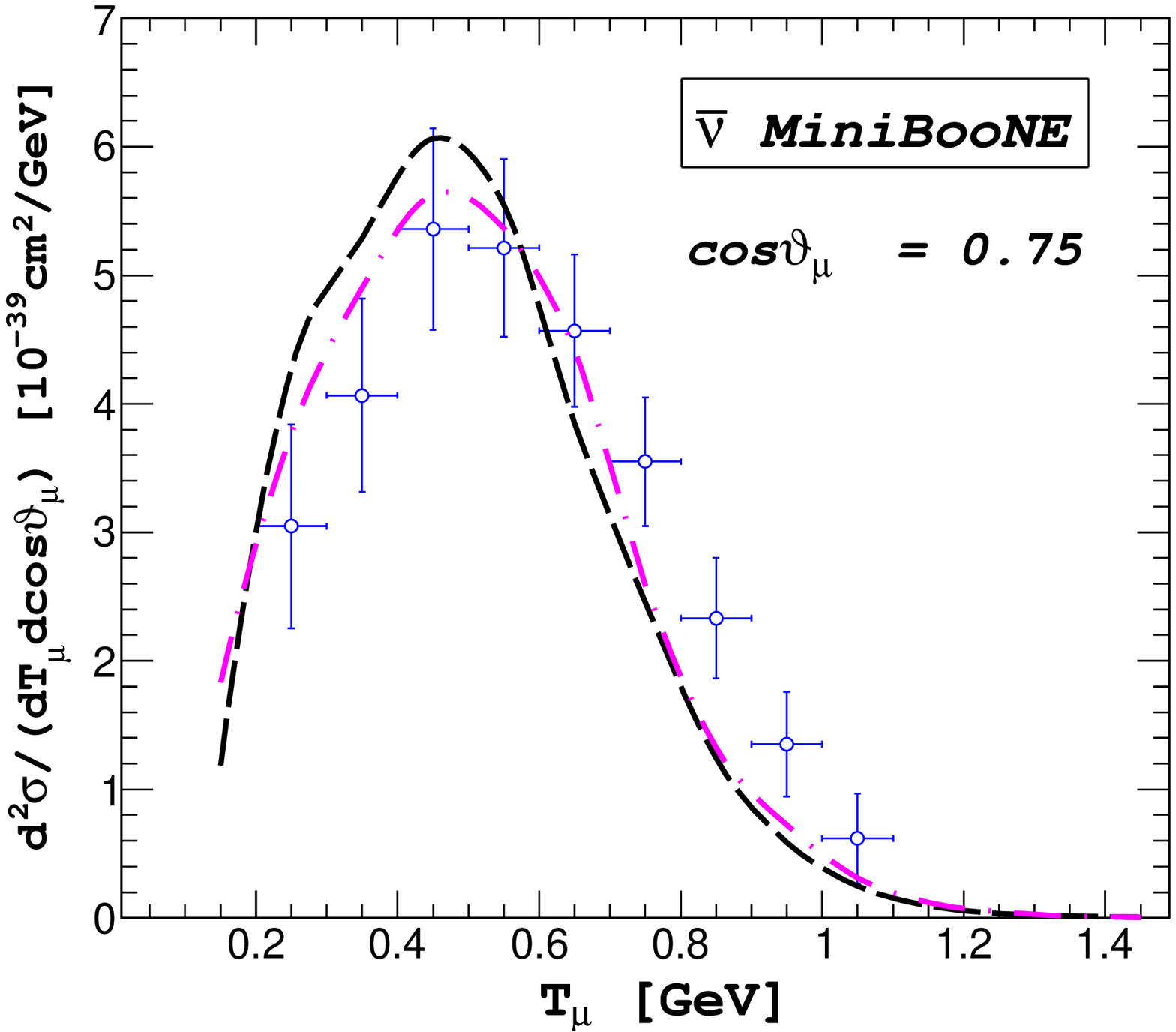}
\caption{Flux-averaged double-differential cross section per 
target nucleon for the CCQE $^{12}$C$(\nu_{\mu} , \mu ^-)$ (left panel) and 
$^{12}$C$(\bar\nu_{\mu} , \mu ^+)$ (right panel) reactions as a function of 
the outgoing muon kinetic energy $T_{\mu}$ for the $\cos\vartheta_{\mu} = 0.75$ 
angular bin. calculated with the RGF-DEM and the RGF-EDAI. 
Experimental data from~\cite{miniboone}.
}
\label{MBCCQE} 
\end{figure}

An example is  presented in Figure~\ref{MBCCQE}, where the RGF 
double-differential $\nu_{\mu}$ and  $\bar{\nu_{\mu}}$  cross sections 
averaged over the MiniBooNE fluxes are displayed
as a function of the muon kinetic energy $T_\mu$ for the 
 $\cos\vartheta_{\mu} = 0.75$ angular bin.
In the calculations the bound nucleon states 
are self-consistent Dirac-Hartree solutions derived within a relativistic
mean-field approach~\cite{Serot1986}. The results of two different 
parameterizations of the ROP are compared: the energy-dependent and 
A-dependent democratic (DEM)~\cite{chc1} 
and the energy-dependent but A-independent EDAI~\cite{chc}. While DEM is a 
global parameterization, which depends on the atomic number $A$ and is obtained 
through a fit to more than 200 data sets of elastic proton-nucleus scattering 
data on a wide range of nuclei,  EDAI  is a single-nucleus parameterization, 
which is constructed to better reproduce the elastic proton-$^{12}$C 
phenomenology.  
In Figure~\ref{MBCCQE} both RGF-DEM and RGF-EDAI are in reasonable agreement 
with the data around the peak region, while the data are slightly 
underpredicted for large  $T_{\mu}$. 
 
\begin{figure}[t]
\centering
\includegraphics[scale=0.27]{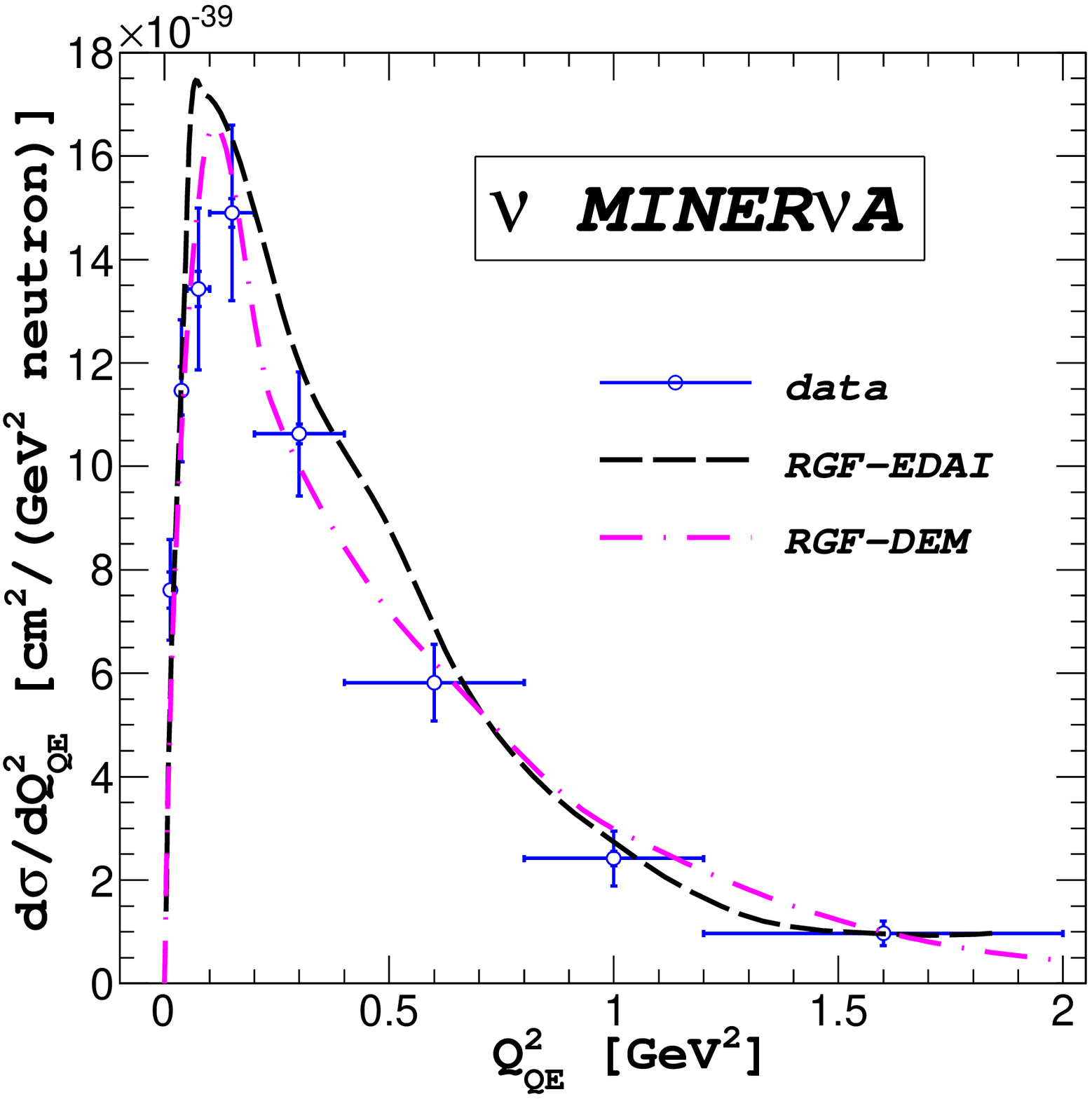} 
\includegraphics[scale=0.27]{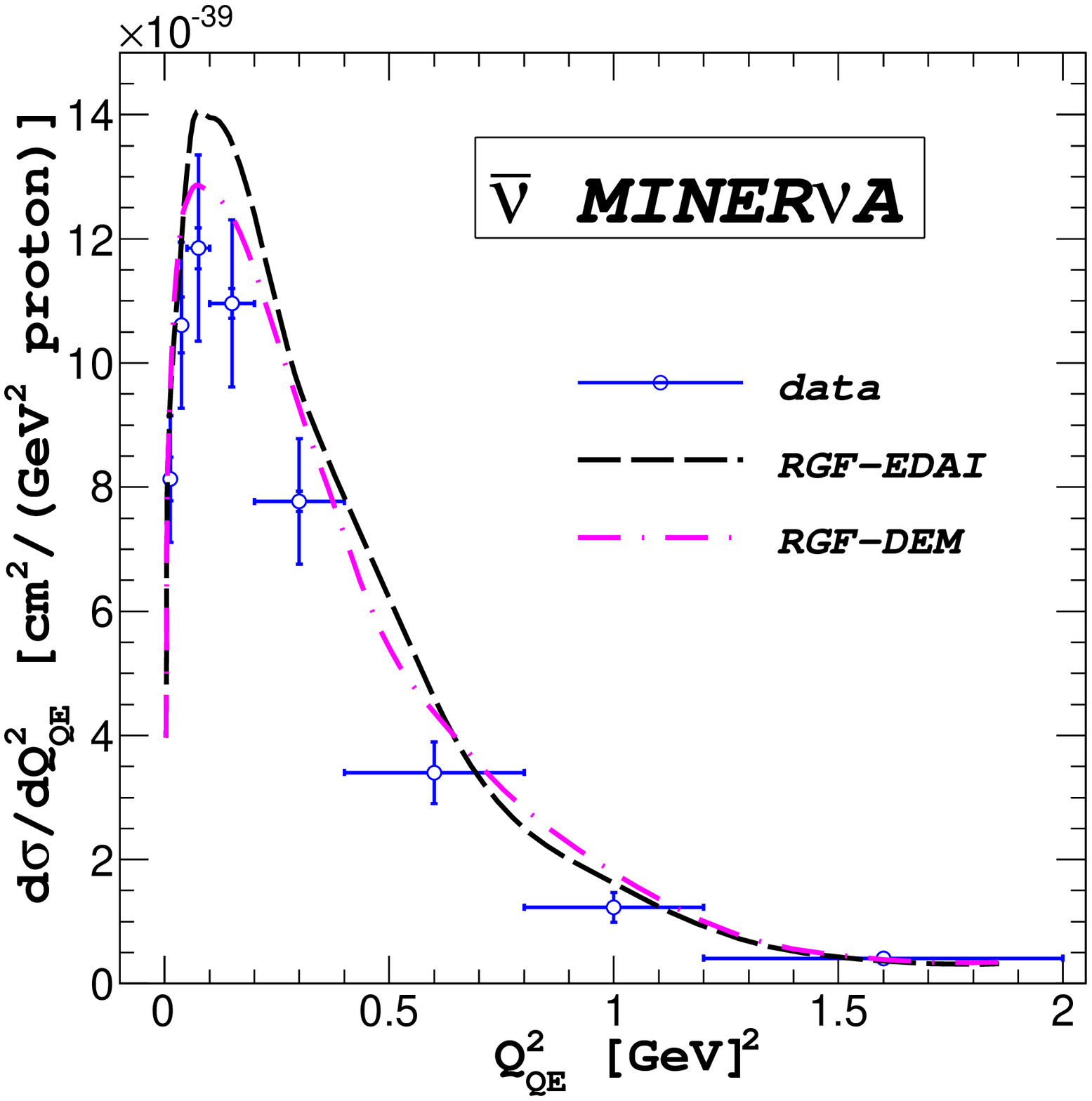}
\caption{Flux-averaged differential $\nu - ^{12}$C (left panel) and 
$\bar{\nu} - ^{12}$C (right panel) cross section per 
target nucleon as a function of of $Q^2_{QE}$ calculated with the RGF-DEM and 
the RGF-EDAI. Experimental data from~\cite{minerva1,minerva2}.
}
\label{minerva} 
\end{figure}

The MINER$\nu$A  Collaboration has recently measured differential cross 
sections for $\nu$ and $\bar\nu$ CCQE scattering on a hydrocarbon 
target~\cite{minerva1,minerva2}. The experimental energy 
is higher than MiniBooNE. The RMF, which undepredicts the MiniBooNE data, provides 
a good description of the CCQE MINER$\nu$A data~\cite{juan-minerva}. The RGF 
results give a satisfactory description of the MiniBooNE CCQE data and are very sensitive to 
the kinematic conditions of the calculations.
In Figure~\ref{minerva}  the RGF-EDAI and RGF-DEM differential flux
averaged cross sections $d\sigma/dQ^{2}_{QE}$ for $\nu$  and $\bar\nu$ 
scattering off a CH target are displayed as a function of the reconstructed 
four-momentum transfer squared $Q^{2}_{QE}$ and compared with the 
MINER$\nu$A data. 
Both  RGF-EDAI and RGF-DEM are in good agreement with the data. 
The RGF-EDAI cross sections are, however, larger than 
the RGF-DEM ones for $Q^{2}_{QE}\lesssim 0.5$ GeV$^2$, while similar results 
are obtained with the two ROPs for larger values of $Q^{2}_{QE}$.

At MiniBooNE kinematics the RGF results are significantly larger than 
the RMF ones and in better agreement with the MiniBooNE 
data~\cite{compmini,prd}. The differences are reduced at MINER$\nu$A kinematics: 
the RGF results in Fig.~\ref{minerva} are still somewhat larger 
than the RMF ones of~\cite{juan-minerva} but in agreement with the data within 
the experimental errors.

\begin{figure}[t]
\centering
\includegraphics[scale=0.27]{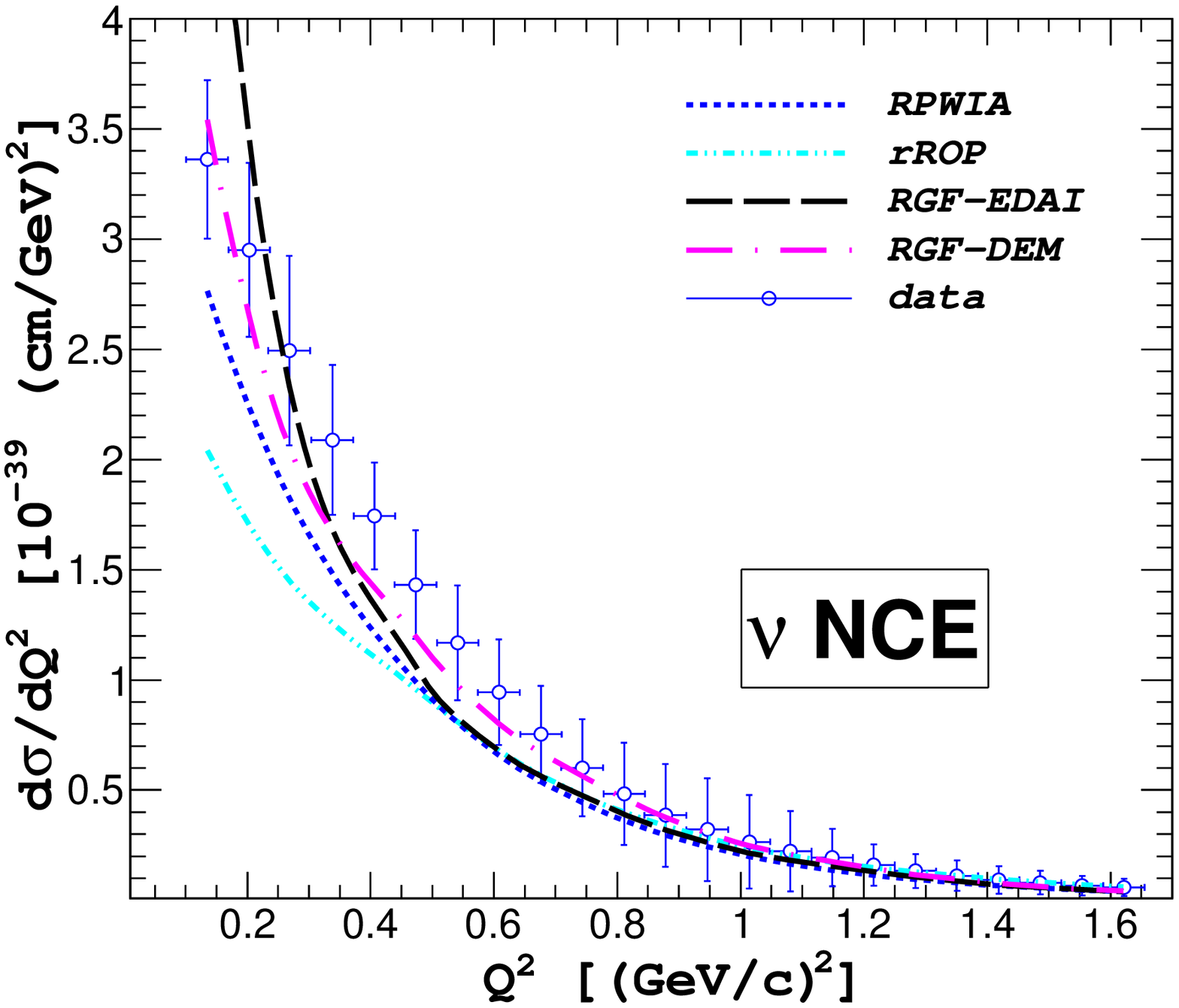} 
\includegraphics[scale=0.27]{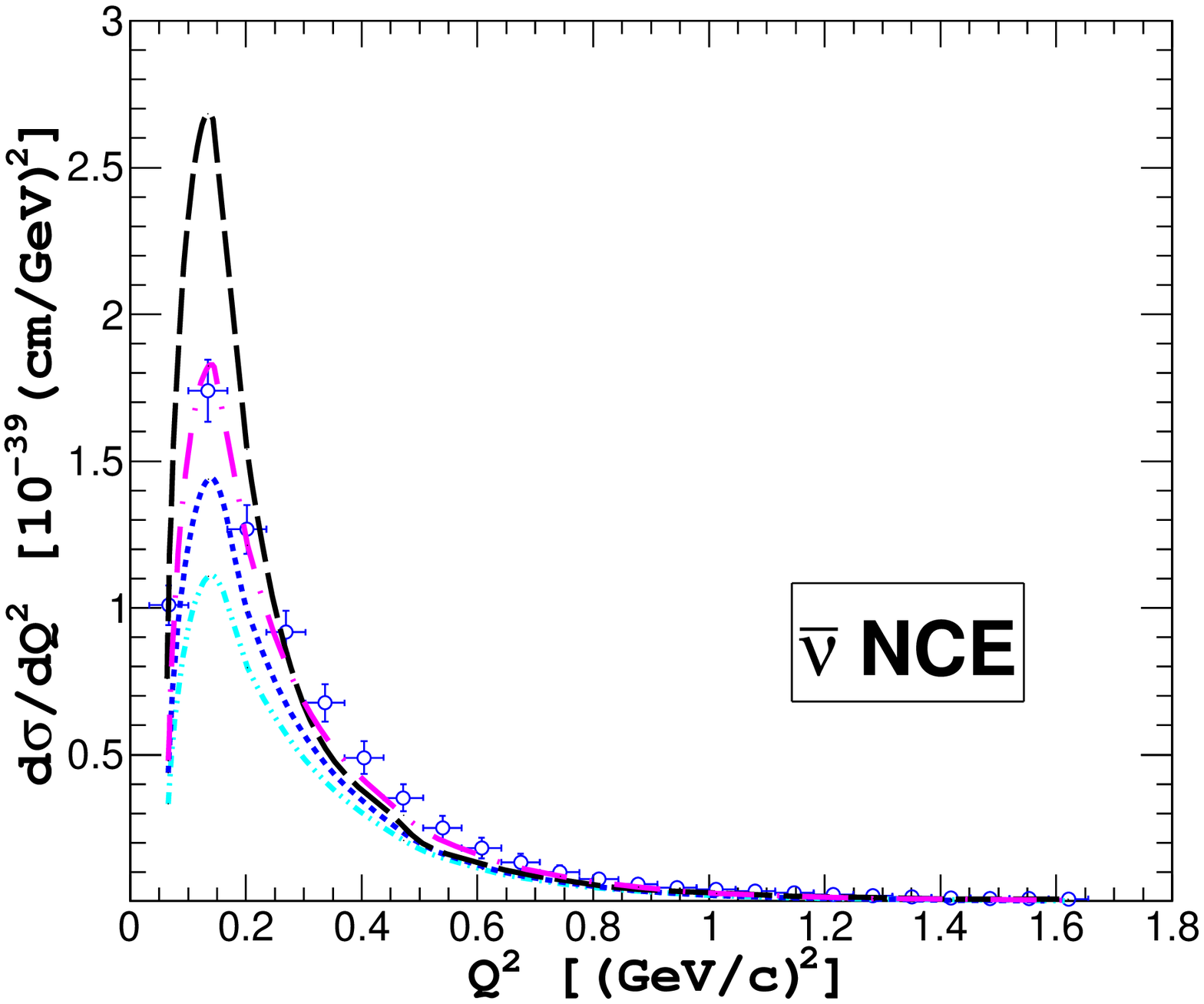}
\caption{NCE flux-averaged  differential $(\nu N \rightarrow \nu N)$ 
(left panel) and  $(\bar{\nu} N \rightarrow \bar{\nu} N)$ (right panel) 
cross sections as a function of $Q^2$ calculated with the RPWIA, rROP, RGF-DEM,
RGF-EDAD1, and the RGF-EDAI. Experimental data 
from~\cite{miniboonenc, miniboonencan}.
}
\label{nce} 
\end{figure}
\begin{figure}[t]
\centering
\includegraphics[scale=0.29]{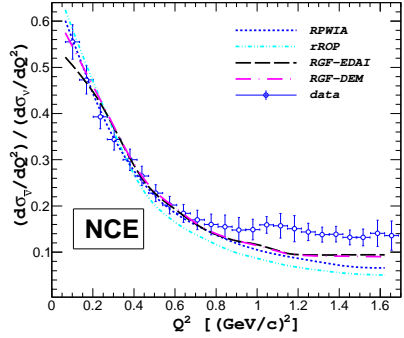} 
\caption{Ratio of the $\bar{\nu}$ to $\nu$ NCE scattering cross sections with 
total errors. Experimental data 
from~\cite{miniboonencan}.
}
\label{rationce} 
\end{figure}

The MiniBooNE Collaboration has also measured  NCE flux-averaged differential
$\nu$~\cite{miniboonenc} and $\bar\nu$~\cite{miniboonencan} cross 
sections on CH$_2$ as a function of $Q^2$. 
The analysis of NCE reactions introduces additional 
complications, as the final neutrino cannot be measured and a final hadron has 
to be detected. Thus, NCE cross sections are usually
semi-inclusive in the hadronic sector,
where events for which at least one nucleon in the final state
is detected are experimentally selected.
The description of the semi-inclusive NCE scattering with the RGF can
recover important contributions that are not present in the RDWIA, 
which is appropriate for the exclusive scattering but neglects some  
channels which can contribute to the semi-inclusive reaction. The RGF, however, 
describes the inclusive process and, as such, it may include channels which are 
not present in the semi-inclusive NCE measurements.  From this point of view, 
the RDWIA can represent a lower limit and the RGF an upper limit to the 
NCE cross sections.
In comparison with the MiniBooNE NCE data, the RDWIA generally underpredicts
the experimental cross section, while the RGF results are in reasonable
agreement with the NCE data~\cite{prd}.

In Fig.~\ref{nce},  the 
RPWIA, rROP, and  RGF $(\nu N \rightarrow \nu N)$
and $(\bar{\nu} N \rightarrow \bar{\nu} N)$ cross sections are displayed as 
a function of $Q^2=2m_N \sum_i T_i$. The variable $Q^2$ is defined 
assuming that the target nucleon is at rest, $m_N$ being the nucleon mass and 
$\sum_i T_i$ the total kinetic energy of the outgoing nucleons.
The RPWIA results, where FSI are neglected, show a satisfactory, although not 
perfect, agreement with the magnitude of the data whereas
the rROP ones, which are  calculated retaining 
only the real part of the EDAI potential, underestimate the data but for 
$Q^2 \geq 0.6$ (GeV$/c)^2$. The RMF cross
sections in~\cite{compnc} are lower than the rROP ones and therefore below the data.
Larger cross sections and a better agreement with the NCE data is provided by 
the RGF. The differences between RGF-DEM and RGF-EDAI calculations are due to 
the different imaginary parts of the two  ROPs. 
The real parts are similar and give in practice the same rROP results. 
The best agreement with the data is provided by the RGF-DEM results. 

In Fig.~\ref{rationce}, the ratio of the $\bar{\nu}$ to $\nu$ NCE cross sections are
also presented.  The MiniBooNE Collaboration performed both $\nu$
and $\bar{\nu}$ NCE measurements in the same beam line
and with the same detector but with opposite horn polarities and, despite 
the fact that the experimental $\nu$
and $\bar{\nu}$ fluxes are not identical, the ratio of the two cross sections
should minimize the errors and provide a useful observable
to test various theoretical models.
In Fig.~\ref{rationce} all the models give very
close results. In particular, the RGF results are practically
independent of the parameterization adopted for the ROP. All the results
are within the experimental errors, but fom large $Q^2$, where
they slightly underestimate the data.

\section{Conclusions}

We have reviewed some recent results of the RGF model.
The model was developed to describe FSI in the inclusive QE 
electron scattering, it has been tested in comparison with electron-scattering 
data, and it has then been extended to  QE neutrino scattering. 

The RGF results are usually larger than the results of other models based 
on the RIA and are able to describe the CCQE MiniBooNE and MINER$\nu$A data 
and the NCE MiniBooNE data, both for $\nu$ and $\bar\nu$ scattering.

The RGF model is based on the use of a complex energy-dependent ROP whose 
imaginary part includes the overall effect of the inelastic channels, which 
give different contributions at different energies and make the RGF results 
sensitive to the kinematic conditions of the calculations. 
The use of the complex ROP allows us to include all the available
final-state channels and not only direct one-nucleon emission processes.
The important role of contributions other than direct one-nucleon emission has 
been confirmed by different and independent models in the case of
MiniBooNE cross sections, but  the same  conclusion is doubtful in the case of 
the MINER$\nu$A data.

The RGF  does not include two-body currents, but it can include rescattering 
processes,  non-nucleonic $\Delta$ excitations, 
and also some multinucleon processes. 
Although  not incorporated explicitly in the model, part of these contributions 
can be recovered, to some extent, by the imaginary part of the ROP. 
With the use of phenomenological ROPs, however, we cannot disentangle 
different reaction processes and explain in detail the origin of the enhancement. 

The numerical predictions of the RGF are affected by the theoretical uncertainties 
in the determination of the phenomenological ROP. A better determination 
which closely fulfills the dispersion relations 
deserves further investigation.

\end{document}